\documentclass[conference]{IEEEtran}
\IEEEoverridecommandlockouts
\usepackage{cite}
\usepackage{amsmath,amssymb,amsfonts}
\usepackage{algorithmic}
\usepackage{graphicx}
\usepackage{textcomp}
\usepackage{xcolor}

\usepackage{multirow}
\usepackage{makecell}
\usepackage{pbox}
\usepackage{breakurl}
\usepackage{url}

\def\BibTeX{{\rm B\kern-.05em{\sc i\kern-.025em b}\kern-.08em
    T\kern-.1667em\lower.7ex\hbox{E}\kern-.125emX}}
        
\usepackage{fancyhdr}
\newcommand{\smallemojiwidth}{0.02\textwidth}

\begin{document}


\title{EMMA: An Emotion-Aware Wellbeing Chatbot}

\author{\IEEEauthorblockN{Asma Ghandeharioun}
\IEEEauthorblockA{\textit{MIT Media Lab}\\
Cambridge, MA, US \\
asma\_gh@mit.edu}
\and
\IEEEauthorblockN{Daniel McDuff}
\IEEEauthorblockA{\textit{Microsoft Research}\\
Redmond, WA, US \\
damcduff@microsoft.com}
\and
\IEEEauthorblockN{Mary Czerwinski}
\IEEEauthorblockA{\textit{Microsoft Research}\\
Redmond, WA, US \\
marycz@microsoft.com}
\and
\IEEEauthorblockN{Kael Rowan}
\IEEEauthorblockA{\textit{Microsoft Research}\\
Redmond, WA, US \\
kael.rowan@microsoft.com}
}

\maketitle

\begin{abstract}
The delivery of mental health interventions via ubiquitous devices has shown much promise. A conversational chatbot is a promising oracle for delivering appropriate just-in-time interventions. However, designing emotionally-aware agents, specially in this context, is under-explored. Furthermore, the feasibility of automating the delivery of just-in-time mHealth interventions via such an agent has not been fully studied. In this paper, we present the design and evaluation of EMMA (EMotion-Aware mHealth Agent) through a two-week long human-subject experiment with N=39 participants. EMMA provides emotionally appropriate micro-activities in an empathetic manner. We show that the system can be extended to detect a user's mood purely from smartphone sensor data. Our results show that our personalized machine learning model was perceived as likable via self-reports of emotion from users. Finally, we provide a set of guidelines for the design of emotion-aware bots for mHealth.
\end{abstract}


\section{Introduction}

We increasingly rely on intelligent agents in our everyday lives. For these systems to be trusted, natural and engaging, they need to be able to have emotional intelligence. An assistant that can sense a user's emotional state and therefore, adapt, is considered more valuable, intelligent and trustworthy~\cite{bickmore2001relational,gratch2007creating,lucas2014s}. Virtual agents have shown success in multiple contexts, including intelligent tutoring systems~\cite{d2007toward}, health care decision support~\cite{devault2014simsensei}, and more recently as virtual therapists~\cite{ring2016affectively}.

Advances in affective computing~\cite{picard1997affective} over the past twenty years mean that it is now possible to deploy applications in-situ and longitudinally. Computer sensing platforms can now track a user's state across time~\cite{mcduff2012affectaura}, which presents the opportunity to personalize interactions with individuals based on their affective state. Not only desktop computers, but also smartphones and wearable devices have been studied to conduct ``Reality Mining" \cite{eagle2006reality} and to infer the user's context and mood \cite{likamwa2013moodscope,  likamwa2011can}.

A very promising application for intelligent agents is in the delivery of mental health therapies. Prior work has shown that simple micro-interventions, such as deep breathing or talking with a friend~\cite{paredes2014poptherapy} or practicing an act of kindness \cite{ghandeharioun2016kind} can be effective in increasing positive affect and reducing negative affect.  Mobile mental health is of growing interest, as it leverages ubiquitous devices and can be used to reach people, regardless of their location.  Furthermore, smartphones and watches are equipped with a wide variety of sensors that can be very useful in affect detection. However, the affective qualities of an agent delivering such an intervention are poorly understood. Is it beneficial if the agent expresses emotion? Can an agent learn to react emotionally appropriately given the context and user? Does an emotionally intelligent agent magnify the impact of an intervention? 

In the area of mental health, there are still open questions about how to use technology to sense affective states and, more importantly, how to effectively provide interventions should one need help. Might recipients be more receptive to technologies that are more affectively neutral, resulting in the technology being trusted more or considered more objectively intelligent? Or should designers try to resemble a counselor or trusted companion, designing for a more empathic and human experience during a technological intervention?

In this paper, we introduce the design of EMMA (EMotion-Aware mHealth Agent), an emotionally intelligent wellness personal assistant for the general population. EMMA provides relevant micro-activities for mental wellness in an empathetic manner and learns to detect mood from smartphone location data. We evaluate different aspects of EMMA through a two-week long human-subject experiment with N=39 participants. This experiment is a randomized trial, comparing two groups: EMMA, and a control condition. This experiment explores the introduction of machine learning (ML) models for automating affect detection and its influence on users' perception of the system. The first week was focused on capturing training data and the models were deployed during the second week. Our results showed that the chatbot that automated mood detection using personalization and location data from the phone was perceived equally as likable as the bot relying on one's self-reported emotion samples. We further explored the influence of EMMA on latency and frequency of response to interventions. 

\section{Related Work}\label{sec:relatedWork}

Despite multiple attempts by several researchers, classifying subjective metrics related to wellbeing and mood remains a difficult task, with relatively low accuracies, ranging from 55\% to 80\%. Examples include using smartphone data to model social interactions \cite{dong2011modeling}, to study the relationship between mood and sleep \cite{moturu2011using}, to detect stress, happiness, and mood \cite{sano2013stress, bogomolov2014daily, bauer2012can,likamwa2013moodscope,bogomolov2013happiness, jaques2015predicting}, and to predict depressive symptoms \cite{saeb2015mobile}. Others have also attempted prediction of fine grained symptoms on a continuous scale using smartphone data and wearable sensors \cite{ghandeharioun2017objective}. Though not perfect, personal sensing -``collection and analysis of data from sensors embedded in the context of daily life with the aim of identifying human behaviors, thoughts, feelings, and traits" ~\cite{mohr2017personal} - has shown potential for monitoring mental health and providing just-in-time interventions.

Ecological momentary interventions (EMIs) are becoming more popular, especially for the treatment of clinical depression and anxiety. They have been effective at reducing symptoms of depression and anxiety, reducing outcomes of stress, and increasing positive psychological functioning \cite{schueller2017ecological}. Automated text-messaging, used as an adjunct to therapy, has helped users stay in therapy for longer, and attend more sessions \cite{aguilera2017automated}. Synchronous, text-based interventions, either by a human or a chat-bot, have shown significant mental health outcome improvements compared to a wait-list condition \cite{hoermann2017application}. 

There are endless subtleties in designing automated text interventions for mental health purposes. Tailoring \cite{muench2017randomized} and diversifying \cite{kocielnik2017send} messages have shown potential for improving efficacy and reducing habituation. Sender, stimulus type, delivery medium, heterogeneity, timing of delivery, frequency, intensity, the trigger's target, structure, narrative \cite{muench2017more}, and the linguistic content of messages \cite{muench2014understanding} are among the variables that need to be optimized for the purpose of the intervention. Other researchers have addressed low engagement and high attrition in self-guided web-based interventions by building a peer support platform - Panoply \cite{morris2015efficacy, morris2015crowdsourcing} - and using a conversational agent - woebot ~\cite{fitzpatrick2017delivering}.

Conversational agents have shown promise in automating the detection of psychological symptoms for both assessment and the evaluation of treatment impact ~\cite{miner2016conversational}.
 There is evidence suggesting that the general population can also benefit from such eHealth interventions. Anxiety and depression prevention EMIs are associated with small but positive effects on symptom reduction. The medium to long-term effects of such interventions need further exploration~\cite{deady2017ehealth}.

In positive computing \cite{calvo2014positive} literature, there have been efforts around personalizing interventions toward the users' preferences (e.g.,  \cite{jeong2016improving, paredes2014poptherapy}) and using sensor data to derive the timing of interventions (e.g., \cite{ghandeharioun2016kind, sano2017designing}). Moreover, conversational agents that are emotionally expressive have shown promise for behavior change applications \cite{ghandeharioun2019towards}. However, targeting relevant micro-activities toward a full range of emotional states, varying the tone of delivery appropriately, and exploring automation feasibility has not been fully studied.

\section{Method}\label{sec:systemDesign}

\begin{figure}[h!]
	\centering
  	\includegraphics[width=0.425\columnwidth]{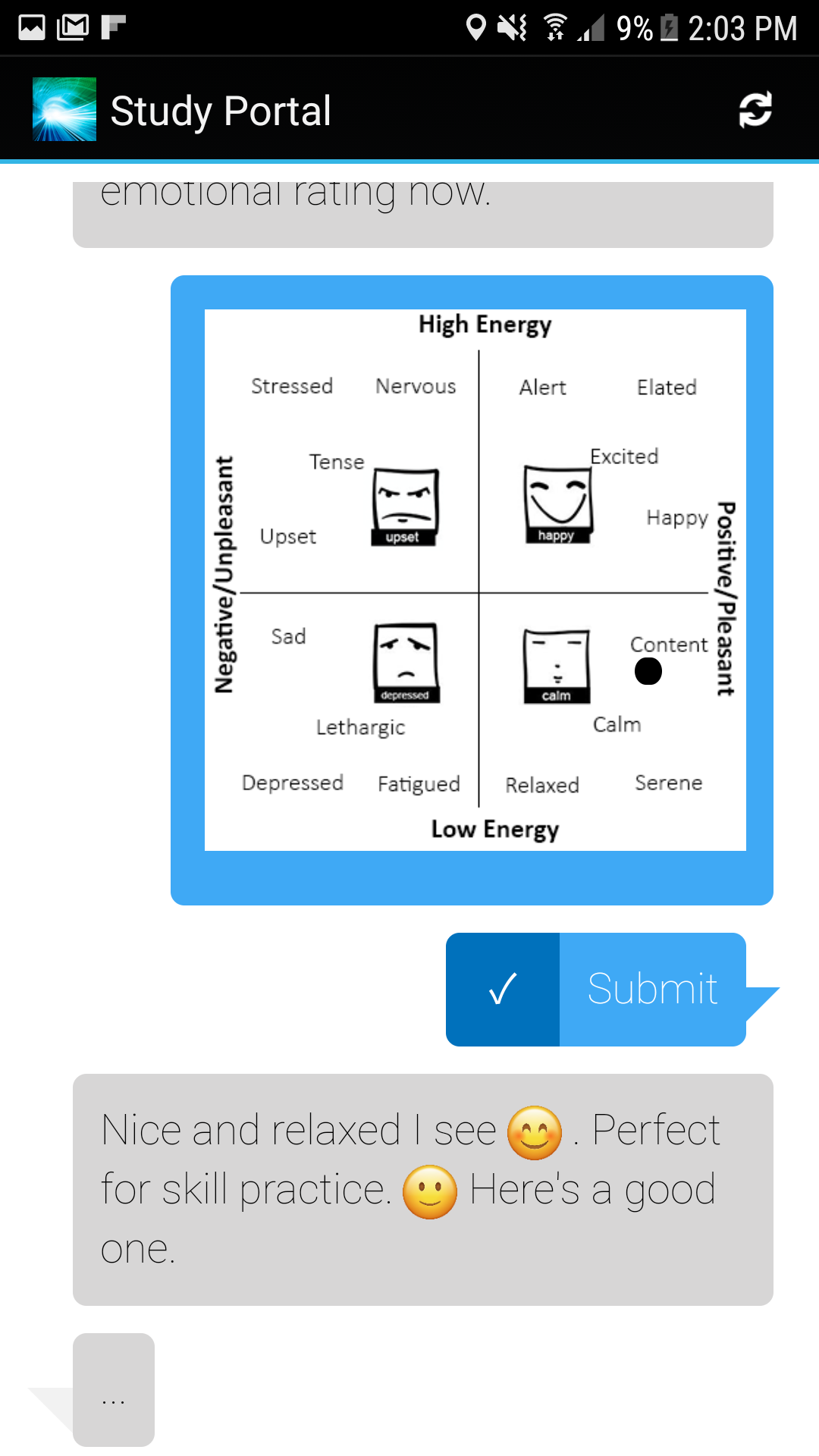}
  	\caption{The visual design of the EMMA user interface.}
    	\label{fig:UI}
        \vspace{-0.5cm}
 \end{figure}
 
EMMA is an extension to an emotion-aware experience sampling chatbot that we built \cite{ghandeharioun2019towards}. In this section, we describe how we extend the mobile app to measure phone sensors, use ML to infer mood from sensor data, suggest appropriate wellness activities, and seamlessly put them all into context with affective surrounding text and adjust the app's behavior based on group condition and study's temporal phase (Fig. \ref{fig:UI}). 





\subsection{Inferring Affect}
We continuously captured geolocation and detailed activities within the application to get contextual information from the phone\footnote{Accelerometer data, calls and messages metadata, and calendar events were also captured. However, due to the high missing data rate, we decided to solely focus on location data. The missing data were due to differences in the availability of sensor data on different versions of the Android OS.}. To preserve battery power while capturing location, we set the movement threshold to 10 meters and uploaded the captured location once every minute. We were able to capture at least 50 location data points from 97\% of the participants, including 294279 total location data points. The loggers captured data periodically in the foreground and background.

We translated the raw data into higher level features for each hour. Our features included average latitude, average longitude, standard deviation of latitude, and standard deviation of longitude during every hour. We also included average distance from work. Since all participants were internal members of the same institution, the work location was approximated by the building's latitude and longitude. We also included distance from home, where home was approximated by the median of the location when the user was not at work. We also encoded time of the day and day of the week as contextual information. These types of location features have precedent in prior mHealth studies{~\cite{saeb2015mobile}}. Additionally, personal measures from pre-study surveys were included: user ID, gender, baseline scores of the big five personality test \cite{digman1990personality}, PANAS (Positive and Negative Affect Scale, short version) \cite{watson1988development}, and DASS (Depression, Anxiety and Stress Scale) \cite{lovibond1995structure}. PANAS quantifies mood and DASS captures depression, anxiety, and stress symptoms. For categorical variables such as user ID and gender, we used their one-hot representation: when a variable has $d$ distinct possible values, it substitutes each observation with $d$ binary values, indicating the presence (1) or absence (0) of the $d$th value. 
The prediction engine, explained in Section~\ref{sec:ML}, uses these features to infer mood.

Additionally, to capture ground-truth emotion labels, we administered experience sampling five times a day using a visual grid (Fig. \ref{fig:UI}) based on Russel's two-dimensional model of emotion \cite{russell1980circumplex}. Note that self-reports were only used to validate automatic sensor-based predictions of mood.

\subsection{Wellbeing Interventions}
We built upon previous work on micro-interventions for improving wellness \cite{paredes2014poptherapy, sano2015healthaware, deady2017ehealth}. This set of interventions includes individual or social short activities that fall into one of the following psychotherapy categories: positive psychology, cognitive behavioral, meta-cognitive, or somatic interventions. The activities provide a textual prompt and a link to an online tool for executing the activity. This set of interventions has shown reduction in depressive symptoms and improvement in stress coping capabilities over the course of 4 weeks \cite{paredes2014poptherapy}.

We revisited these activities to make them more appropriate for different emotional states. We have assigned each micro-activity to the most relevant quadrant(s) on the 2x2 Russell circumplex model of emotion  \cite{russell1980circumplex}. The interventions were augmented to have 16 activities per quadrant. Table \ref{tab:affectiveInterventions} shows a sample intervention for each quadrant.

\begin{table}[h!]
\caption{An example of wellbeing interventions targeted at emotional states. TL, TR, BL and BR refer to the spatial locations on the 2x2 circumplex model of emotion, e.g. TL: Top Left quadrant.}
   \centering
   \begin{tabular}{ p{0.5cm}  p{7.5cm} }
     \textbf{State} & \textbf{Sample Intervention} \\
     TL & Write yourself a note with some issue that could wait for longer. \\
     TR & Spread the joy by calling a friend and passing along your positive energy! \\
     BL & Affirmations always make us feel better. Check some of these out and share them with some friends. \\
     BR & Celebrate with others! Write a positive comment to some friend's good posting. \\
   \end{tabular}
     \label{tab:affectiveInterventions}
 \end{table}
\vspace{-.5cm}

\subsection{Emotionally Expressive Delivery}
We have scripted different emotionally charged phrasings for each possible interaction and randomly selected one when communicating with the user. For example, if the user was classified in the BL quadrant of Russel's circumplex model of emotion, the chatbot would recommend an activity by saying: ``Feeling glum? \includegraphics[width = \smallemojiwidth]{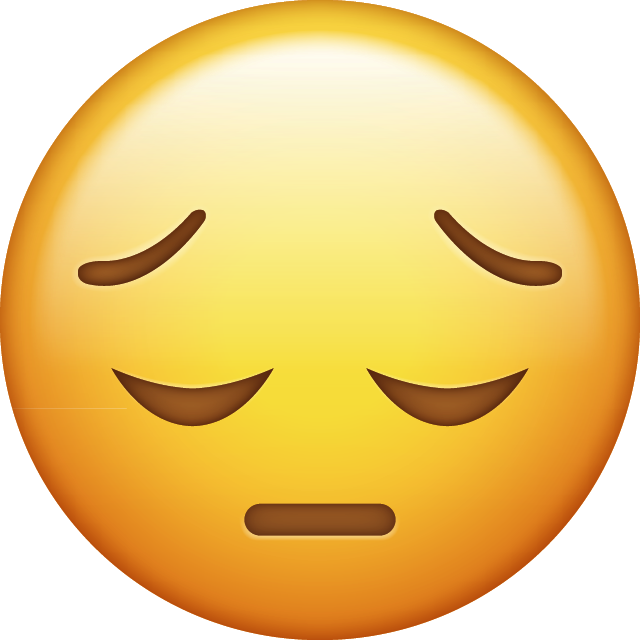} I have a skill that might brighten your day. \includegraphics[width = \smallemojiwidth]{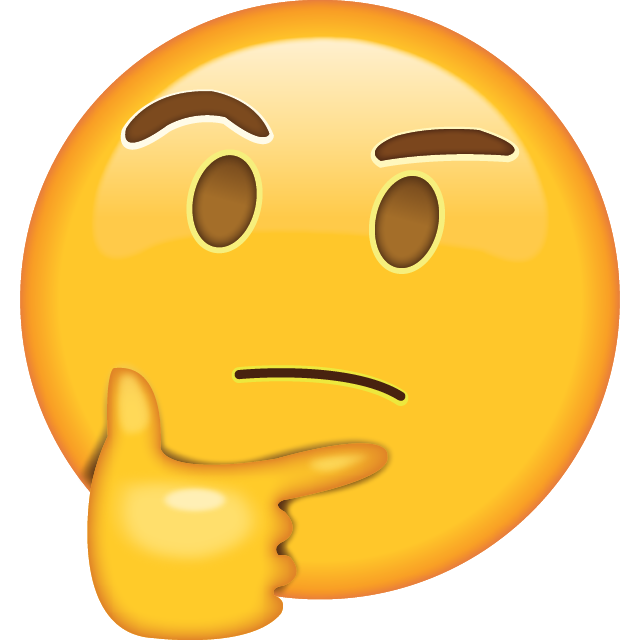} Let's practice.''. For the control condition, we scripted similar texts, but without any expression of affect or use of emojis. For example, the parallel to the above example would be: ``Okay. Let's try an intervention then."  


\section{Human Subjects}\label{sec:humanSubjects}

\begin{table}[h!]
   \centering
   \begin{tabular}{ l  c  c c  c c c }
     \textbf{Group} & \textbf{Total} & \multicolumn{2}{c}{\textbf{Gender}} & \multicolumn{3}{c}{\textbf{Employment}} \\
      & & Female & Male & FTE & Intern & Other\\
     EMMA & 19 & 3 & 16 & 8 & 9 & 2\\
     Control & 20 & 4 & 16 & 9& 8 & 3\\
   \end{tabular}
   \caption{Participation demographics per group.}
     \label{tab:groups}
 \end{table}
\vspace{-0.5cm}

The study protocol was approved by the institutional review board at Microsoft. Table \ref{tab:groups} summarizes the group assignments and demographics of participants. The population was generally mentally healthy\footnote{Baseline DASS \cite{lovibond1995structure} scores were captured. Mean values were within suggested normal ranges, i.e. below 4.5 for depression scale, below 3.5 for anxiety scale, and below 7 for stress scale. }. Gift-card raffles were held at the end of each week, for \$75, and \$100 respectively. Three participants were randomly selected as winners of each raffle\footnote{All participants were part of a bigger project and received \$200 upon successfully completing all studies.}. 

\section{Experiment: Intervention Effectiveness, Scalability, and Automation}\label{sec:experiment2}

Our first research question is regarding the capacity to scale and automate the bot so that it predicts emotion labels only from the user's phone usage behavior and does not require constant self-report of emotion labels. This question should be first addressed objectively by calculating the accuracy of mood prediction from phone sensor data. However, it is also important to analyze users' preference to understand if substituting ground-truth emotion labels with a ML prediction influences the likability of the system.

Our second research question is regarding how intervention engagement is mediated by the emotional intelligence of the bot delivering it. Previously researchers have studied response time to phone notifications and accounted perceived disruption as an influencing factor on response time \cite{mehrotra2016my}. Thus, we measure response latency as a proxy for intervention disruption vs. engagement.  We also measure frequency of response to interventions as another engagement quantification metric. 

To answer these questions, we designed a two-week longitudinal experiment. We randomized participants into two groups: EMMA, and Control. During the first week, the EMMA group had access to the mobile app that administered experience sampling, detected user's selected emotional quadrant, and responded with emotionally relevant phrases. In addition, EMMA would randomly select from a set of interventions that were emotionally appropriate for the user's current state. EMMA would deliver the intervention surrounded with emotionally expressive text, scripted for that quadrant. The Control group received a similar experience, in terms of triggering experience sampling and providing emotionally relevant interventions; however, the bot was not emotionally expressive itself. Though it understood which quadrant has been selected by the user and provided skills accordingly, all the surrounding text was neutral, without any expression of emotion.

During the second week, a ML model simultaneously predicted the user's current affect. This prediction was the basis of the suggested intervention in both EMMA and Control conditions. In the EMMA condition, the surrounding affectively expressive text was also driven by the prediction. The self-reported emotion labels were still being stored on the cloud, but only used later as the ground-truth measure for calculating accuracy of the ML emotion detection model. Below, we explain the ML model selection, training, and validation.

\subsection{Machine Learning Models}
\label{sec:ML}
To translate the sensor data into affect, we developed a prediction engine. We used the data from all but last week of the experiment, and split it into train and test sets (75\% and 25\% of samples respectively). We trained multiple models on the training set, used 10-fold cross validation for parameter optimization within each model category, and used the hold-out test set for selecting the best model for the second week of the experiment. Our criteria for best model selection were performance, simplicity, and explainability, in that order. We also report a baseline where the classifier always predicts the most frequent class in the training set. Specifically for unbalanced data, this is stronger than a random chance classifier.

\subsubsection{Classification Models}
We first implemented binary classifiers for valence (negative/positive) and arousal (low/high) separately. We experimented with a range of classifiers including Logistic Regression, Ridge, AdaBoost, Bagging, Random Forest, and Gaussian Processes.

\subsubsection{Regression Models}
Additionally, we tried modeling valence and arousal on a continuous scale. We normalized the valence and arousal values and experimented with a range of regression models including Linear Regression, several regularized versions of linear regression (Ridge, Lasso, Elastic Net), Bayesian Ridge, Support Vector Regression, Gradient Boosting, AdaBoost, Random Forest, and robust to outlier methods (RANSAC, Theil-Sen, and Huber). We later quantized the predicted values to calculate accuracy measures.

\subsubsection{Personalized Regression Models}
Individuals tend to have different baselines and oscillate around those values. To better model such personal patterns, we calculated the average of valence ($v_b$) and arousal ($a_b$) in the training set per individual. Then, we explicitly modeled the variation of valence and arousal from $v_b$ and $a_b$, respectively, on a continuous scale using regression models.

In Section \ref{sec:ML-validation}, we show the boost in performance, especially for arousal detection, using personalization. Ultimately, we selected the personalized model with Random Forest regression for valence prediction and AdaBoost regression for arousal prediction, and this is explained in the results section\footnote{Although the final aim is to perform a classification task, what makes the regression model better suit our problem is our ability to predict explicit deviation from personal baseline rather than predicting the absolute value in the label space. A continuous label space would easily allow such transformation while it is not be feasible in a binary label space. We believe that is why the personalized model, although not directly optimizing for classification, works better than the classification models.}.


 \begin{table*}[!t]
 \caption{Training and Validation Phase: Results of the best performing models on the hold-out set. Acc. refers to the accuracy of the model. Model parameters: $e$ - the number of estimators, $c$ - criterion, $m$ - maximum samples, and $\lambda$ - learning rate.}
   \centering
   \begin{tabular}{ p{2.5cm}  p{4cm} p{1cm}  p{4cm} p{1cm}  p{1cm} }
     \multirow{ 2}{*}{\textbf{Method}} & \multicolumn{2}{c}{\textbf{Valence}} & \multicolumn{2}{c}{\textbf{Arousal}} & \textbf{Quadrants} \\
      & \textbf{} & \textbf{Acc.} & \textbf{}  & \textbf{Acc.} & \textbf{Acc.}\\
     \textbf{Classification} & Random Forest$_{(e=10, c=gini)}$ & 80.4\% & Bagging$_{(e=10, m=1.0)}$ & 49.4\% & 41.9\%\\
     \textbf{Regression} & Random Forest$_{(e=10, c=gini)}$ & 80.6\% &  Random Forest$_{(e=10, c=gini)}$ & 50.4\% & 40.1\%\\
     \textbf{Personalized} & Random Forest$_{(e=10, c=gini)}$ & 82.4\% & Ada Boost$_{(e=50, \lambda=1.0)}$ & 67.0\% & 56.8\%\\
     \textbf{Baseline} & Most frequent & 80.6\% & Most frequent & 51.9\% & 42.4\%\\
   \end{tabular}
   \vspace{-.3cm}
     \label{tab:models}
 \end{table*}

\subsubsection{Validation}
\label{sec:ML-validation}
Table \ref{tab:models} summarizes the performance of classification, regression, and personalized regression models on the hold-out set. As expected, the personalized regression model outperformed the classification, non-personalized regression model and the baseline; thus, the personalized regression model was selected for the second week of the experiment. For valence prediction we used the Random Forest and for arousal prediction we used the AdaBoost. As shown in the table, predicting arousal has been more difficult than valence. This could be due to the fact that most participant tended to stay in the same binary valence state, while their arousal value was closer to the neutral condition and bounced more frequently between low and high energy.

To further confirm the performance of the selected model, personalized regression, we calculated Pearson correlation coefficients between the predicted and actual values for the hold-out test-set. There was a significant correlation between predicted and actual arousal (r=.43, p$\ll$.0001, n=387), and a significant correlation between predicted and actual valence (r=.57, p$\ll$.0001, n=387).

\subsection{Measures}

\subsubsection{Latency in Response to Interventions}
To test our hypotheses regarding the interplay between emotional intelligence of the bot and intervention engagement, we captured and analyzed the latency in response to interventions. We define response latency as the time between receiving a notification and responding to it in minutes. This measure is extracted from the application logs of user clicks in the app UI.

\subsubsection{Frequency of Response to Interventions}
We extract response frequency as the average number of responses to interventions per participant, per week, from the app usage logs. This measure is a surrogate for intervention engagement.

\subsubsection{User Preference}
We assessed satisfaction and efficacy of the system through different questions using a Likert scale, ranging from 1 (strongly disagree) to 7 (strongly agree). These questions asked about agent's likability, intelligence, and appropriateness of its ``tone". We asked about user preference for continuing to interact with the agent, and his/her improvement in awareness of daily emotions. We also asked if the notifications from the app where too frequent. Also, we included an open-ended question for general comments. This measure was captured at the end of each week. The questions are provided in the Supplementary Materials section.

\subsubsection{Experience Sampling}
Using the visual experience sampling grid, we captured valence ($v$) and arousal ($a$) on a continuous scale, $v, a\in [0.0,1.0]$. We used a 0.5 threshold to discretize $v$ into $\hat v$ which encodes positive vs. negative valence. We discretized $a$ similarly to derive $\hat a$ which encodes high vs. low arousal. We used binary values of $\hat v$ and $\hat a$ for calculating accuracy of our ML models on valence and arousal separately. The 4 possible combinations of $(\hat v,\hat a$) are mapped to the 4 quadrants on the visual grid: Top Left (TL), Top Right (TR), Bottom Left (BL), and Bottom Right (BR). We used quadrant accuracy for selecting the best performing ML model.

\subsection{Results}

\subsubsection{Quantitative Performance}

  \begin{table*}[!t]
  \caption{Test Phase. Results of deployment (final week). Acc. - accuracy, $e$ - the number of estimators, $c$ - criterion, $\lambda$ - learning rate.}
   \centering
   \begin{tabular}{ p{2.5cm}  p{4cm} p{1cm}  p{4cm} p{1cm}  p{1cm} }
     \multirow{ 2}{*}{\textbf{Method}} & \multicolumn{2}{c}{\textbf{Valence}} & \multicolumn{2}{c}{\textbf{Arousal}} & \textbf{Quadrants} \\
      & \textbf{Best model} & \textbf{Acc.} & \textbf{Best model}  & \textbf{Acc.} & \textbf{Acc.}\\
     \textbf{Personalized} & Random Forest$_{(e=10, c=gini)}$ & 82.2\% & Ada Boost$_{(e=50, \lambda = 1.0)}$ & 65.7\% & 56.6\%\\
     \textbf{Baseline} & Most frequent & 82.3\% & Most frequent & 48.0\% & 41.5\%\\
   \end{tabular}
   \vspace{-0.3cm}
     \label{tab:w3model}
 \end{table*}

After deploying the personalized regression model in the second week of the experiment, we did similar post-hoc analyses to calculate objective performance of the model. Table \ref{tab:w3model} summarizes the results.

We also calculated Pearson correlation coefficients between the predicted and actual values for the final week. There was a significant correlation between predicted and actual arousal (r=.54, p$\ll$.0001, n=702), and a significant correlation between predicted and actual valence (r=.43, p$\ll$.0001, n=702).

\subsubsection{User Perception}

The objective performance measures show that the model had reasonable accuracy during the automation phase. But did the users agree? Did they find the first week of the experiment that used ground-truth emotion samples as likable as the second week that used ML predictions? Or did the occasional prediction errors reduce the perceived likability of the agent significantly? To answer this question, we compared the self-reported agent evaluation for when it was driven by ML vs. experience sampling.

We employed two one-sided t-tests (TOST) as a test for non-inferiority on the average of all likability measures before and after deploying ML. We set the equivalence intervals as follows: $\Delta L = \Delta U = 0.5$. We tested the two resulting composite null hypotheses: $H01: \Delta \le -\Delta L$ and $H02: \Delta \ge \Delta U$. The results were $t(38)=5.31$, $p\ll0.0001$ and $t(38)=-6.33$, $p\ll0.0001$, respectively. Since both of these one-sided tests are statistically rejected, we conclude that the likability of the agent is practically equivalent before and after deploying ML and there is no significant decline in overall agent preference as measured by the average of all the likability measures, though no improvement either. This is a promising result, suggesting that ML models could provide a scalable affect-driven agent that does not require constant user effort for providing self-reports, and users perceive it just as favorably.

\subsubsection{Intervention Engagement}

\begin{figure}[t!]
\centering
  	\includegraphics[width=0.47\columnwidth]{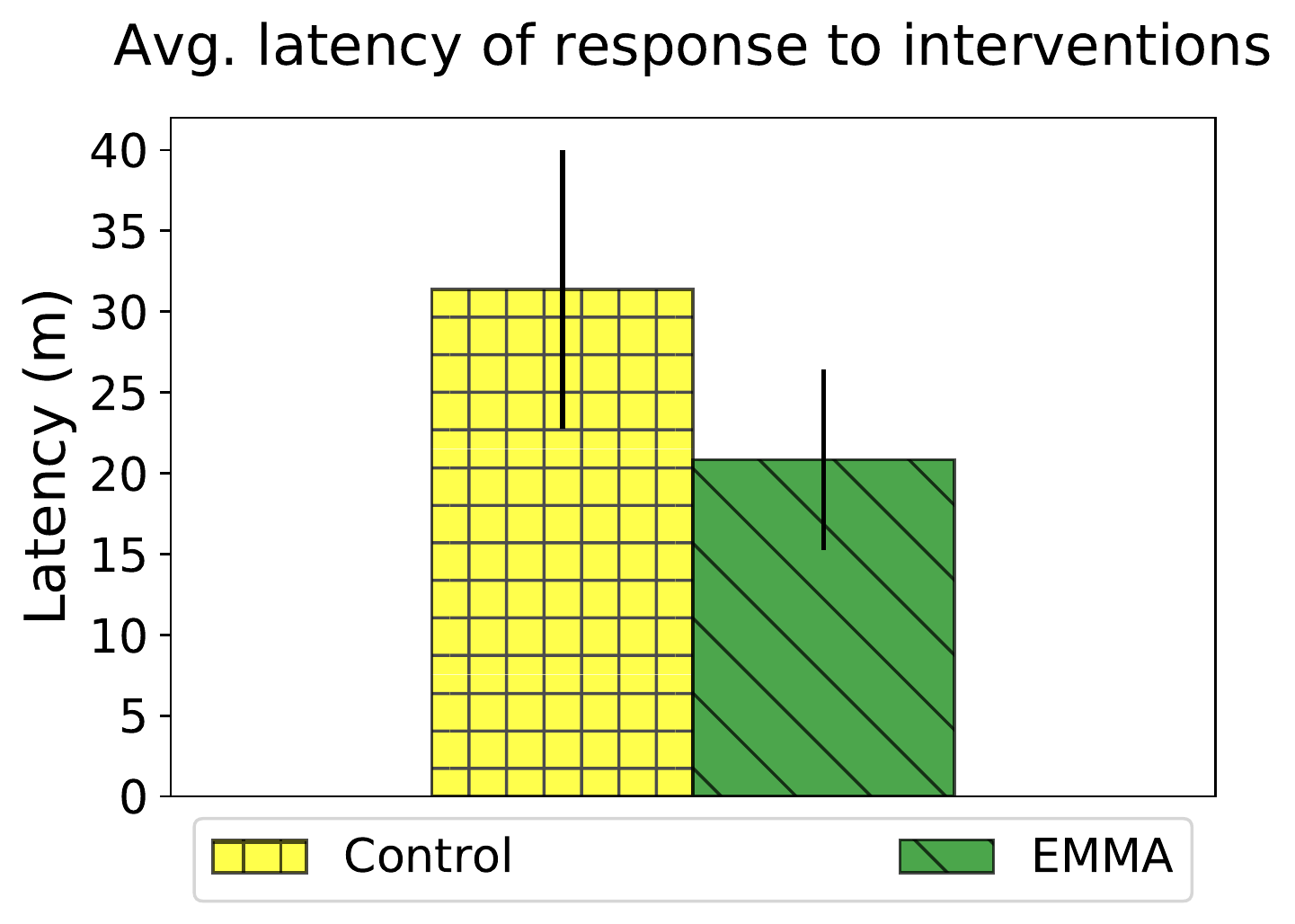}
    \includegraphics[width=0.5\columnwidth]{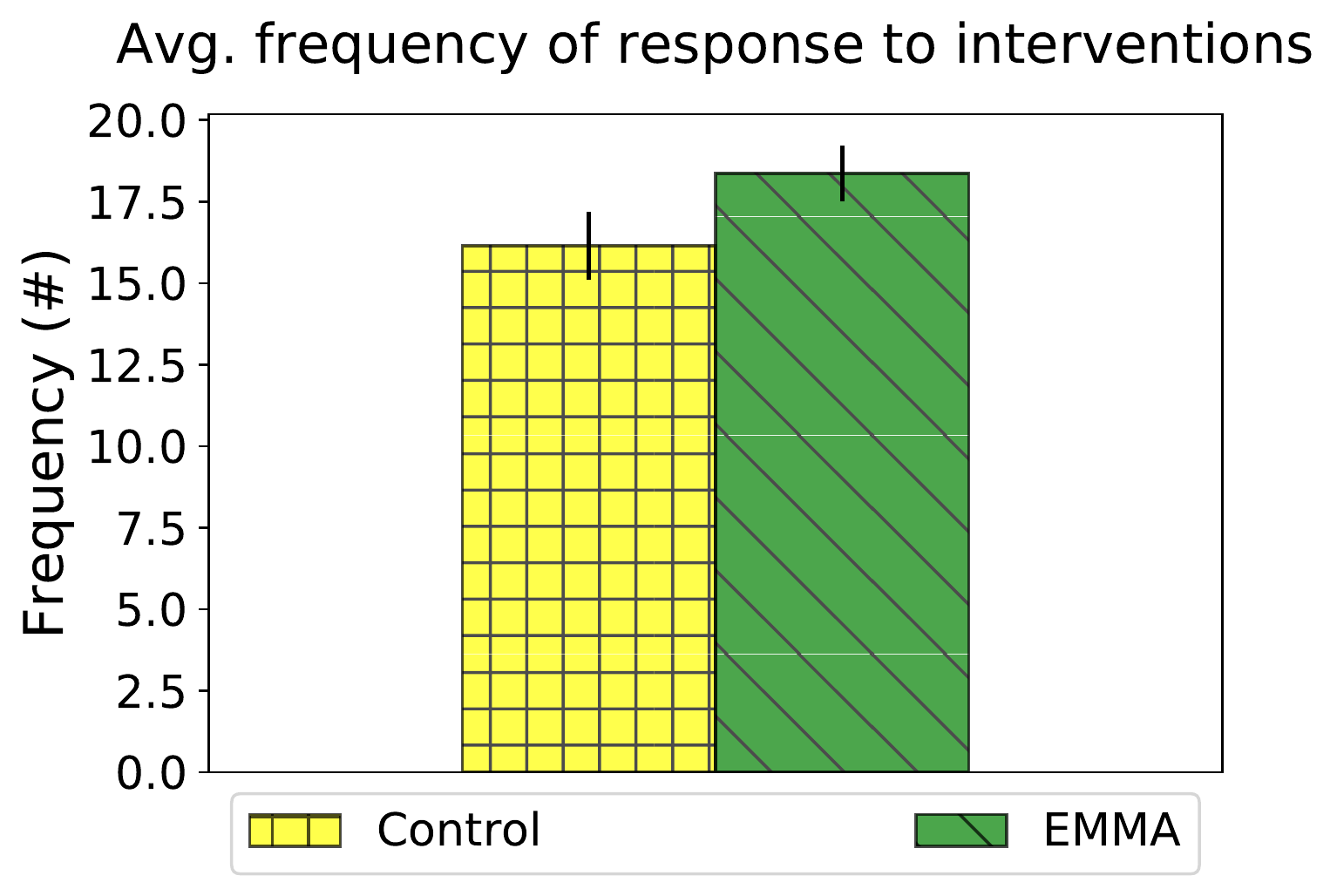}
  \caption{Response latency and frequency with one standard error bars.}
   \label{fig:latencyFrequencyW2}
\end{figure}

Fig. \ref{fig:latencyFrequencyW2} visualizes the intervention response time and frequency for each group. We observed a trend suggesting that participants in the EMMA condition tended to respond more quickly and to a higher number of interventions compared to the control group. However, an independent t-test between EMMA and the control condition did not reach statistical significance at .05 level for response latency or frequency\footnote{$t_{l}(37)=-.99, p=.32; t_{f}(37)=1.59, p=.11$. Future studies are needed for further validation.}.


\subsubsection{Qualitative Feedback}

Some users mentioned enjoying interacting with the app; pa041: ``I love being part of this study. The app is great, the surveys are short, and it's been fun thinking about my emotions."; pa052: ``I did find it interesting to use the app and become aware of how stable my emotions are. That was the most positive outcome for me in this study.'' 

Responses showed individual differences among users' preferences about interventions, however. Most users preferred shorter and simpler activities; pa063: ``The most successful activities have involved watching short videos or images."; pa067: ``I preferred the interventions that I could do on the phone without making any noise."; pa064: ``Simple things, like \textit{do a stretch} or \textit{read a joke} or \textit{think about this kind of fond memory} were generally helpful."

Some participants mentioned that the activities were not always optimized for the context, they did not have time for them, or they did not like them. These points were brought up by users from all groups. For example, pa035: ``I'm frequently in the middle of other things when the notification shows up and I don't have time or it's inappropriate for me to engage with my phone for 5-10 minutes.''; pa038: ``it doesn't take busyness into account."; pa040: ``It has suggested that I walk over to a colleague's office; but I was working remotely so that wasn't possible."; pa041: ``They seem like fantastic suggestions. I'm just not going to stop what I'm doing."; pa064: ``I found it very difficult to engage with many of the skills that agent presented to me, due to time, the local environment I was in, or lack of interest." pa057: Some of the tasks we were asked to do were not applying to me. For example I have not posted anything on Facebook and I was uncomfortable posting some random stuffs after a while."

Importantly, several participants mentioned they preferred not to be interrupted when feeling positive; pa081: ``If someone indicates that they are feeling happy and/or positive, they shouldn't have to do an activity."; pa077: ``I find it annoying that when I report myself as \textit{happy} or \textit{content}, it still has exercises for me, that typically end up making my mood less positive."; pa080: ``I felt that when I reported positive emotional state it shouldn't then try and improve my mood further with an exercise. I am already feeling positive so an intervention will just distract me and lower my mood."

Some participants mentioned the tone of the agent had become expected, and thus not as effective; pa040: ``The first couple of times I saw feedback on my ratings it was kind of neat; but now it just feels like it is expected that the app will tell me this, so it doesn't really have an effect on me."
 This suggests that personalizing the feedback from the agent based on the context and preferences of the user would be preferable to a rules-based approach as was implemented based on self-reports. As participants started to anthropomorphize EMMA, they expected more richness and variability in their interactions which is in-line with previous research findings \cite{paredes2014poptherapy}.

Some participants mentioned the way the activities were provided sounded prescriptive; pa041: ``I have a hard time giving over control to any kind of app[...]"; pa064: ``the agent should frame the skill as something I can do \textit{if I want to}.''

\section{Discussion}\label{sec:discussion}

\subsection{Automating Affect Detection in an Affective Bot}
We showed that our mobile bot was perceived equally as likable as a bot that works with ground-truth emotion labels captured by experience sampling. This is an encouraging result, as it relies only on smartphone location data, a ubiquitous technology that can significantly reduce the  users' burden of self-reporting during intervention applications. It suggests that automatic - albeit error-prone - affect detection can still be as effective as self-report in certain contexts. 

\subsection{Tailoring Wellness Suggestion Activities to Affective States}
We expected positive states to be good times for practicing skills and building resilience. Also, we expected negative states to benefit more from immediate intervention as a treatment. However, from user feedback we learned that suggesting such activities when a user is in a high energy and positive valence state may have an opposite effect. Note that we focused on a general population rather than clinically depressed individuals. It might be that our healthy participants did not feel the need to practice such skills and found them simplistic, and thus were sometimes annoyed by them. This irritation may have undermined the benefits of practicing such activities in bottom left or top left quadrants of Russel's circumplex model and the role of emotional vs. non-emotional conditions.

\subsection{Guidelines for Affective Chatbot Design and EMI Delivery}

\textit{Do not interrupt a good mood for an EMI.}
Participants mentioned the high rate of interruption by personal technological devices and not wanting to be controlled by them for unnecessary reasons. Our population expressed that when they were in a high energy and positive valence mood, they were already engaged in rewarding activities; thus, interrupting them for an intervention would annoy them and sometimes resulted in a less positive mood. However, they found the activities more useful when in a low energy and negative valence mood.

\textit{Short, simple, and effortless activities are better received.}
Participants mentioned that they were more likely to perform shorter and simpler activities. This highlights the fact that success of an activity in a self-guided mHealth setting first depends on how likely it is to be performed. This calls for the design of more effortless interventions such as \cite{ghandeharioun2017brightbeat}.

\textit{Contextual relevance makes EMIs more respectful.}
Users' feedback revealed that making EMIs contextually relevant is one of the most important elements in designing an intelligent system. The simplest way to mitigate this is to ask participants upfront what times they would like to receive triggers. Taking into account busyness, time of the day, and sensor data to detect context switching are other ways to optimize timing of triggers. This is in line with previous findings (e.g. \cite{paredes2014poptherapy,ghandeharioun2016kind}).

\textit{Diversifying content is required to prevent habituation.}
Habituation is one of the main reasons of interventions being ignored. Starting with a big enough pool of interventions can delay habituation. However, more dynamic methods can sustain the system in the long-term. Novel ways of combining exploitation and exploration to maximize efficacy of personalized suggestions \cite{rabbi2015automated}, including ML techniques to automate content creation, and using peer support can be example solutions to this problem \cite{morris2015efficacy, morris2015crowdsourcing}.

\textit{Providing an opt-out choice is needed for a respectful EMI.}
Especially for a population with relatively low scores on depression, anxiety, and stress scales, which do not qualify for clinical depression or anxiety, users may prefer to maintain control over receiving interventions and providing an opt-out choice may be necessary for the EMI system to be perceived as respectful and intelligent--and ultimately, useful. 

\subsection{Limitations}

We relied on the authors' expertise in psychology and affective computing to assign interventions to their appropriate emotional state. Due to the high  missing data rate from multiple potential sources, we were unable to fully capture context. In the future, we would like to evaluate the appropriateness of intervention assignment through a user study and explore more sophisticated ML models to better leverage sparse data.


\section{Conclusions}\label{sec:conclusions}

We present EMMA, the first emotionally-intelligent and expressive mHealth agent, that provides wellness suggestions in the form of micro-interventions. We quantitatively and qualitatively evaluated EMMA in a human-subject experiment over the course of 2 weeks, with N=39 participants.

We have shown that our system can detect a user's mood from passive smartphone sensor data and that using automatically predicted emotional states to drive emotional dialog and the choice of interventions did not impact people's opinions of the agent versus manual EMI entry.  This finding means we could reduce the burden on the user to report their emotions and make EMMA highly scalable. 

Our longitudinal study allowed us to identify several design guidelines for future work. Specifically, we found that delivering interventions was not effective for those people already in a high activation positive mood, and that diversity of dialog and content is necessary to avoid habituation. Our observations highlighted the importance of contextual relevance, simplicity, and reserving an opt-out choice for successful EMIs. We believe that, if interventions are more focused to specific moods and contexts, and are personalized and less predictable, they have the potential to improve positive affect.

\vspace{12pt}

\bibliographystyle{IEEEtran}
\bibliography{EMMA}

\begin{thebibliography}{10}
\providecommand{\url}[1]{#1}
\csname url@samestyle\endcsname
\providecommand{\newblock}{\relax}
\providecommand{\bibinfo}[2]{#2}
\providecommand{\BIBentrySTDinterwordspacing}{\spaceskip=0pt\relax}
\providecommand{\BIBentryALTinterwordstretchfactor}{4}
\providecommand{\BIBentryALTinterwordspacing}{\spaceskip=\fontdimen2\font plus
\BIBentryALTinterwordstretchfactor\fontdimen3\font minus
  \fontdimen4\font\relax}
\providecommand{\BIBforeignlanguage}[2]{{%
\expandafter\ifx\csname l@#1\endcsname\relax
\typeout{** WARNING: IEEEtran.bst: No hyphenation pattern has been}%
\typeout{** loaded for the language `#1'. Using the pattern for}%
\typeout{** the default language instead.}%
\else
\language=\csname l@#1\endcsname
\fi
#2}}
\providecommand{\BIBdecl}{\relax}
\BIBdecl

\bibitem{bickmore2001relational}
T.~Bickmore and J.~Cassell, ``Relational agents: a model and implementation of
  building user trust,'' in \emph{CHI}.\hskip 1em plus 0.5em minus 0.4em\relax
  ACM, 2001, pp. 396--403.

\bibitem{gratch2007creating}
J.~Gratch, N.~Wang, J.~Gerten, E.~Fast, and R.~Duffy, ``Creating rapport with
  virtual agents,'' in \emph{International Workshop on Intelligent Virtual
  Agents}.\hskip 1em plus 0.5em minus 0.4em\relax Springer, 2007, pp. 125--138.

\bibitem{lucas2014s}
G.~M. Lucas, J.~Gratch, A.~King, and L.-P. Morency, ``It’s only a computer:
  Virtual humans increase willingness to disclose,'' \emph{Computers in Human
  Behavior}, vol.~37, pp. 94--100, 2014.

\bibitem{d2007toward}
S.~D'Mello, R.~W. Picard, and A.~Graesser, ``Toward an affect-sensitive
  autotutor,'' \emph{IEEE Intelligent Systems}, vol.~22, no.~4, 2007.

\bibitem{devault2014simsensei}
D.~DeVault, R.~Artstein, G.~Benn, T.~Dey, E.~Fast, A.~Gainer, K.~Georgila,
  J.~Gratch, A.~Hartholt, M.~Lhommet \emph{et~al.}, ``Simsensei kiosk: A
  virtual human interviewer for healthcare decision support,'' in
  \emph{Proceedings of the 2014 international conference on Autonomous agents
  and multi-agent systems}.\hskip 1em plus 0.5em minus 0.4em\relax
  International Foundation for Autonomous Agents and Multiagent Systems, 2014,
  pp. 1061--1068.

\bibitem{ring2016affectively}
L.~Ring, T.~Bickmore, and P.~Pedrelli, ``An affectively aware virtual therapist
  for depression counseling,'' in \emph{Proceedings of the CHI 2016 workshop on
  Computing and Mental Health}, 2016.

\bibitem{picard1997affective}
R.~Picard, \emph{Affective computing}.\hskip 1em plus 0.5em minus 0.4em\relax
  MIT press Cambridge, 1997, vol. 252.

\bibitem{mcduff2012affectaura}
D.~McDuff, A.~Karlson, A.~Kapoor, A.~Roseway, and M.~Czerwinski, ``Affectaura:
  an intelligent system for emotional memory,'' in \emph{CHI}.\hskip 1em plus
  0.5em minus 0.4em\relax ACM, 2012, pp. 849--858.

\bibitem{eagle2006reality}
N.~Eagle and A.~Pentland, ``Reality mining: sensing complex social systems,''
  \emph{Personal and ubiquitous computing}, vol.~10, no.~4, 2006.

\bibitem{likamwa2013moodscope}
R.~LiKamWa, Y.~Liu, N.~D. Lane, and L.~Zhong, ``Moodscope: Building a mood
  sensor from smartphone usage patterns,'' in \emph{Proceeding of the 11th
  annual international conference on Mobile systems, applications, and
  services}.\hskip 1em plus 0.5em minus 0.4em\relax ACM, 2013, pp. 389--402.

\bibitem{likamwa2011can}
------, ``Can your smartphone infer your mood,'' in \emph{PhoneSense workshop},
  2011, pp. 1--5.

\bibitem{paredes2014poptherapy}
P.~Paredes, R.~Gilad-Bachrach, M.~Czerwinski, A.~Roseway, K.~Rowan, and
  J.~Hernandez, ``Poptherapy: Coping with stress through pop-culture,'' in
  \emph{PervasiveHealth}.\hskip 1em plus 0.5em minus 0.4em\relax ICST, 2014,
  pp. 109--117.

\bibitem{ghandeharioun2016kind}
A.~Ghandeharioun, A.~Azaria, S.~Taylor, and R.~W. Picard, ``"kind and
  grateful": a context-sensitive smartphone app utilizing inspirational content
  to promote gratitude,'' \emph{Psychology of well-being}, vol.~6, no.~1, pp.
  1--21, 2016.

\bibitem{dong2011modeling}
W.~Dong, B.~Lepri, and A.~S. Pentland, ``Modeling the co-evolution of behaviors
  and social relationships using mobile phone data,'' in \emph{Proceedings of
  the 10th International Conference on Mobile and Ubiquitous Multimedia}.\hskip
  1em plus 0.5em minus 0.4em\relax ACM, 2011, pp. 134--143.

\bibitem{moturu2011using}
S.~T. Moturu, I.~Khayal, N.~Aharony, W.~Pan, and A.~Pentland, ``Using social
  sensing to understand the links between sleep, mood, and sociability,'' in
  \emph{PASSAT/SocialCom}.\hskip 1em plus 0.5em minus 0.4em\relax IEEE, 2011,
  pp. 208--214.

\bibitem{sano2013stress}
A.~Sano and R.~W. Picard, ``Stress recognition using wearable sensors and
  mobile phones,'' in \emph{ACII}.\hskip 1em plus 0.5em minus 0.4em\relax IEEE,
  2013, pp. 671--676.

\bibitem{bogomolov2014daily}
A.~Bogomolov, B.~Lepri, M.~Ferron, F.~Pianesi, and A.~S. Pentland, ``Daily
  stress recognition from mobile phone data, weather conditions and individual
  traits,'' in \emph{Proceedings of the 22nd ACM international conference on
  Multimedia}.\hskip 1em plus 0.5em minus 0.4em\relax ACM, 2014, pp. 477--486.

\bibitem{bauer2012can}
G.~Bauer and P.~Lukowicz, ``Can smartphones detect stress-related changes in
  the behaviour of individuals?'' in \emph{PERCOM Workshops}.\hskip 1em plus
  0.5em minus 0.4em\relax IEEE, 2012, pp. 423--426.

\bibitem{bogomolov2013happiness}
A.~Bogomolov, B.~Lepri, and F.~Pianesi, ``Happiness recognition from mobile
  phone data,'' in \emph{SocialCom}.\hskip 1em plus 0.5em minus 0.4em\relax
  IEEE, 2013, pp. 790--795.

\bibitem{jaques2015predicting}
N.~Jaques, S.~Taylor, A.~Azaria, A.~Ghandeharioun, A.~Sano, and R.~Picard,
  ``Predicting students' happiness from physiology, phone, mobility, and
  behavioral data,'' in \emph{ACII}.\hskip 1em plus 0.5em minus 0.4em\relax
  IEEE, 2015, pp. 222--228.

\bibitem{saeb2015mobile}
S.~Saeb, M.~Zhang, C.~J. Karr, S.~M. Schueller, M.~E. Corden, K.~P. Kording,
  and D.~C. Mohr, ``Mobile phone sensor correlates of depressive symptom
  severity in daily-life behavior: an exploratory study,'' \emph{JMIR},
  vol.~17, no.~7, 2015.

\bibitem{ghandeharioun2017objective}
A.~Ghandeharioun, S.~Fedor, L.~Sangermano, D.~Ionescu, J.~Alpert, C.~Dale,
  D.~Sontag, and R.~Picard, ``Objective assessment of depressive symptoms with
  machine learning and wearable sensors data,'' in \emph{ACII}.\hskip 1em plus
  0.5em minus 0.4em\relax IEEE, 2017.

\bibitem{mohr2017personal}
D.~C. Mohr, M.~Zhang, and S.~M. Schueller, ``Personal sensing: Understanding
  mental health using ubiquitous sensors and machine learning,'' \emph{Annual
  Review of Clinical Psychology}, vol.~13, pp. 23--47, 2017.

\bibitem{schueller2017ecological}
S.~M. Schueller, A.~Aguilera, and D.~C. Mohr, ``Ecological momentary
  interventions for depression and anxiety,'' \emph{Depression and anxiety},
  vol.~34, no.~6, pp. 540--545, 2017.

\bibitem{aguilera2017automated}
A.~Aguilera, E.~Bruehlman-Senecal, O.~Demasi, and P.~Avila, ``Automated text
  messaging as an adjunct to cognitive behavioral therapy for depression: A
  clinical trial,'' \emph{JMIR}, vol.~19, no.~5, 2017.

\bibitem{hoermann2017application}
S.~Hoermann, K.~L. McCabe, D.~N. Milne, and R.~A. Calvo, ``Application of
  synchronous text-based dialogue systems in mental health interventions:
  Systematic review,'' \emph{JMIR}, vol.~19, no.~8, p. e267, 2017.

\bibitem{muench2017randomized}
F.~Muench, K.~van Stolk-Cooke, A.~Kuerbis, G.~Stadler, A.~Baumel, S.~Shao,
  J.~R. McKay, and J.~Morgenstern, ``A randomized controlled pilot trial of
  different mobile messaging interventions for problem drinking compared to
  weekly drink tracking,'' \emph{PloS one}, vol.~12, no.~2, p. e0167900, 2017.

\bibitem{kocielnik2017send}
R.~Kocielnik and G.~Hsieh, ``Send me a different message: Utilizing cognitive
  space to create engaging message triggers.'' in \emph{CSCW}, 2017.

\bibitem{muench2017more}
F.~Muench and A.~Baumel, ``More than a text message: Dismantling digital
  triggers to curate behavior change in patient-centered health
  interventions,'' \emph{JMIR}, vol.~19, no.~5, p. e147, 2017.

\bibitem{muench2014understanding}
F.~Muench, K.~van Stolk-Cooke, J.~Morgenstern, A.~N. Kuerbis, and K.~Markle,
  ``Understanding messaging preferences to inform development of mobile
  goal-directed behavioral interventions,'' \emph{JMIR}, vol.~16, no.~2, 2014.

\bibitem{morris2015efficacy}
R.~R. Morris, S.~M. Schueller, and R.~W. Picard, ``Efficacy of a web-based,
  crowdsourced peer-to-peer cognitive reappraisal platform for depression:
  Randomized controlled trial,'' \emph{JMIR}, vol.~17, no.~3, 2015.

\bibitem{morris2015crowdsourcing}
R.~R. Morris, ``Crowdsourcing mental health and emotional well-being,'' Ph.D.
  dissertation, Massachusetts Institute of Technology, 2015.

\bibitem{fitzpatrick2017delivering}
K.~K. Fitzpatrick, A.~Darcy, and M.~Vierhile, ``Delivering cognitive behavior
  therapy to young adults with symptoms of depression and anxiety using a fully
  automated conversational agent (woebot): A randomized controlled trial,''
  \emph{JMIR Mental Health}, vol.~4, no.~2, p. e19, 2017.

\bibitem{miner2016conversational}
A.~Miner, A.~Chow, S.~Adler, I.~Zaitsev, P.~Tero, A.~Darcy, and A.~Paepcke,
  ``Conversational agents and mental health: Theory-informed assessment of
  language and affect,'' in \emph{HAI}.\hskip 1em plus 0.5em minus 0.4em\relax
  ACM, 2016, pp. 123--130.

\bibitem{deady2017ehealth}
M.~Deady, I.~Choi, R.~Calvo, N.~Glozier, H.~Christensen, and S.~Harvey,
  ``ehealth interventions for the prevention of depression and anxiety in the
  general population: a systematic review and meta-analysis,'' \emph{BMC
  Psychiatry}, vol.~17, no.~1, p. 310, 2017.

\bibitem{calvo2014positive}
R.~A. Calvo and D.~Peters, \emph{Positive computing: technology for wellbeing
  and human potential}.\hskip 1em plus 0.5em minus 0.4em\relax MIT Press, 2014.

\bibitem{jeong2016improving}
S.~Jeong and C.~L. Breazeal, ``Improving smartphone users' affect and wellbeing
  with personalized positive psychology interventions,'' in \emph{HAI}.\hskip
  1em plus 0.5em minus 0.4em\relax ACM, 2016, pp. 131--137.

\bibitem{sano2017designing}
A.~Sano, P.~Johns, and M.~Czerwinski, ``Designing opportune stress intervention
  delivery timing using multi-modal data,'' in \emph{ACII}.\hskip 1em plus
  0.5em minus 0.4em\relax IEEE, 2017.

\bibitem{ghandeharioun2019towards}
A.~Ghandeharioun, D.~McDuff, M.~Czerwinski, and K.~Rowan, ``Towards
  understanding emotional intelligence for behavior change chatbots,'' in
  \emph{ACII}.\hskip 1em plus 0.5em minus 0.4em\relax IEEE, 2019.

\bibitem{digman1990personality}
J.~M. Digman, ``Personality structure: Emergence of the five-factor model,''
  \emph{Annual review of psychology}, vol.~41, no.~1, pp. 417--440, 1990.

\bibitem{watson1988development}
D.~Watson, L.~A. Clark, and A.~Tellegen, ``Development and validation of brief
  measures of positive and negative affect: the panas scales.'' \emph{Journal
  of personality and social psychology}, vol.~54, no.~6, p. 1063, 1988.

\bibitem{lovibond1995structure}
P.~F. Lovibond and S.~H. Lovibond, ``The structure of negative emotional
  states: Comparison of the depression anxiety stress scales (dass) with the
  beck depression and anxiety inventories,'' \emph{Behaviour research and
  therapy}, vol.~33, no.~3, pp. 335--343, 1995.

\bibitem{russell1980circumplex}
J.~A. Russell, ``A circumplex model of affect,'' \emph{Journal of Personality
  and Social Psychology}, vol.~39, no.~6, pp. 1161--1178, 1980.

\bibitem{sano2015healthaware}
A.~Sano, P.~Johns, and M.~Czerwinski, ``Healthaware: An advice system for
  stress, sleep, diet and exercise,'' in \emph{ACII}.\hskip 1em plus 0.5em
  minus 0.4em\relax IEEE, 2015, pp. 546--552.

\bibitem{mehrotra2016my}
A.~Mehrotra, V.~Pejovic, J.~Vermeulen, R.~Hendley, and M.~Musolesi, ``My phone
  and me: understanding people's receptivity to mobile notifications,'' in
  \emph{CHI}.\hskip 1em plus 0.5em minus 0.4em\relax ACM, 2016, pp. 1021--1032.

\bibitem{ghandeharioun2017brightbeat}
A.~Ghandeharioun and R.~Picard, ``Brightbeat: Effortlessly influencing
  breathing for cultivating calmness and focus,'' in \emph{Ext. Abstracts
  CHI}.\hskip 1em plus 0.5em minus 0.4em\relax ACM, 2017, pp. 1624--1631.

\bibitem{rabbi2015automated}
M.~Rabbi, A.~Pfammatter, M.~Zhang, B.~Spring, and T.~Choudhury, ``Automated
  personalized feedback for physical activity and dietary behavior change with
  mobile phones: a randomized controlled trial on adults,'' \emph{JMIR mHealth
  and uHealth}, vol.~3, no.~2, 2015.

\end{thebibliography}

\end{document}